\documentclass[aps,prl,twocolumn,table,dvipsnames,usenames]{revtex4-2}

\UseRawInputEncoding

\usepackage{graphicx}
\usepackage{amsmath}
\usepackage{amssymb}
\usepackage{float}
\usepackage{chngcntr}

\usepackage{multirow}

\usepackage{titlesec}

\usepackage{enumitem}

\usepackage[colorlinks=true,citecolor=Blue,linkcolor=RubineRed,urlcolor=Blue]{hyperref}
\usepackage{xcolor}

\newcommand{\mbf}{\mathbf}

\begin{document}

\title{Neuromodulation via Krotov-Hopfield Improves Accuracy and Robustness of RBMs}

\author{Ba{\c s}er Tamba{\c s}\href{https://orcid.org/0000-0001-7520-4647}{\includegraphics[scale=0.05]{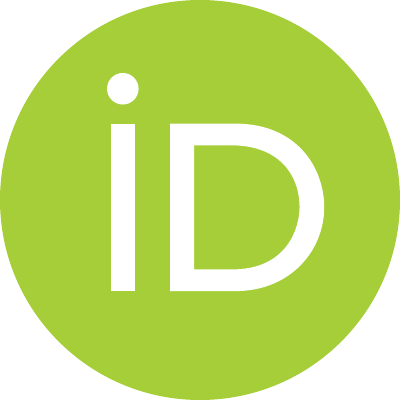}}}
\email[]{tambas19@itu.edu.tr}
\affiliation{Department of Physics, Istanbul Technical University, {\. I}stanbul, T{\" u}rkiye}
\author{A. Levent Suba\c s{\i}\href{https://orcid.org/0000-0001-6489-5665}{\includegraphics[scale=0.05]{orcidid.pdf}}}
\email[]{alsubasi@itu.edu.tr}
\affiliation{Department of Physics, Istanbul Technical University, {\. I}stanbul, T{\" u}rkiye}
\author{Alkan Kabak{\c c}{\i}o{\u g}lu\href{https://orcid.org/0000-0002-9831-3632}{\includegraphics[scale=0.05]{orcidid.pdf}}}
\email[]{akabakcioglu@ku.edu.tr}
\affiliation{Department of Physics, Ko{\c c} University, {\. I}stanbul, T{\" u}rkiye}

\date{\today}

\begin{abstract}
In biological systems, neuromodulation tunes synaptic plasticity based on the internal state of the organism, complementing stimulus-driven Hebbian learning. The algorithm recently proposed by Krotov and Hopfield~\cite{krotov_2019} can be utilized to mirror this process in artificial neural networks, where its built-in intra-layer competition and selective inhibition of synaptic updates offer a cost-effective remedy for the lack of lateral connections through a simplified attention mechanism.  We demonstrate that KH-modulated RBMs outperform standard (shallow) RBMs in both reconstruction and classification tasks, offering a superior trade-off between generalization performance and model size, with the additional benefit of  robustness to weight initialization as well as to overfitting during training.

\end{abstract}

\maketitle

Designing artificial neural networks~(ANNs) that mimic the essentials of biological learning is a challenging task which gained much attention recently~\cite{schmidgall_2024}.
Most ANNs~\cite{lecun_2015} employ the back-propagation algorithm~\cite{rumelhart_1985} which broadcast top-down~(TD) signals through the network for end-to-end weight updates. 
Although highly successful in practice, back-propagation contrasts with the local nature of synaptic plasticity in biological systems.
Recent approaches to bypassing back-propagation include methods using alternative feedback mechanisms---such as feedback alignment~\cite{lillicrap_2016}, target propagation~\cite{lee_2015}, and equilibrium propagation~\cite{scellier_2017}---as well as techniques inspired by Hebbian learning.
Among the latter are Boltzmann machines~\cite{ackley_1985,hinton_2002} (BM), a class of biologically inspired generative neural networks that rely on energy-based dynamics, and more recent methods~\cite{krotov_2019,squadrani_2022} which build on Bienenstock-Cooper-Munro (BCM) theory~\cite{bienenstock_1982}, a refined form of Hebb's rule~\cite{hebb_1949}.

\begin{figure}[t]
\centering
\includegraphics[width=\columnwidth,keepaspectratio]{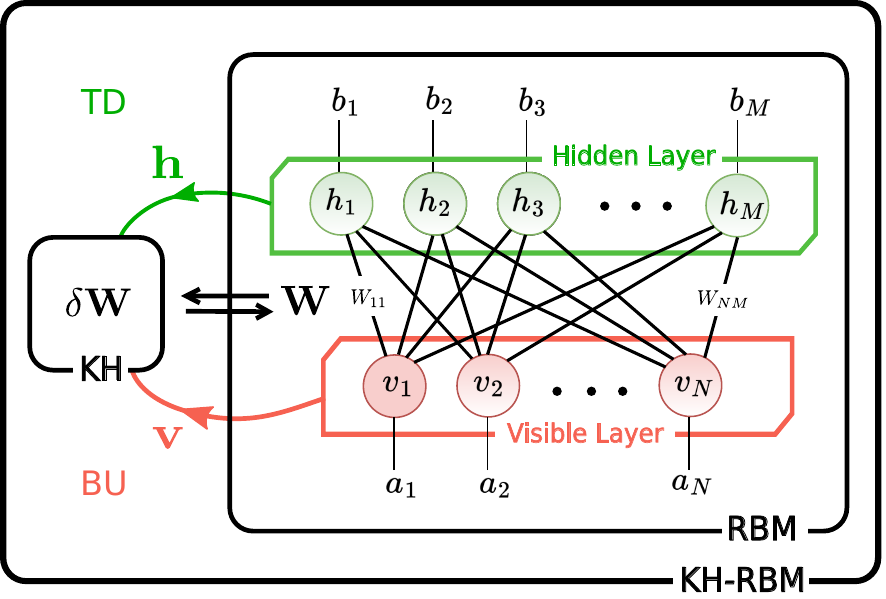}
\caption{\label{fig:graph}Schematic representation of the KH-modulated RBM with 
 top-down (TD) and bottom-up (BU) feedback loops for $\delta \mbf{W}$ weight updates. 
The graph structure of the RBM has weights $\mathbf{W}=\{W_{ij}\}$ linking 
$N$ visible $\mbf{v}=\{v_i\}$ and $M$ hidden $\mbf{h}=\{h_j\}$ 
units with associated bias terms $\mathbf{a}=\{a_i\}$ and $\mathbf{b}=\{b_j\}$, respectively.  
The main text discusses TD feedback only, while results for BU feedback can be found in Supplemental Material Section B.}
\end{figure}

Adoption of BMs in large-scale applications is limited by the intractability of sampling and partition function estimation in fully connected architectures~\cite{salakhutdinov_2009}.
Restricted Boltzmann machines (RBMs) address these challenges by removing lateral connections in the visible and hidden layers (see Fig.~\ref{fig:graph}) which dramatically improves trainability and makes them amenable to analytical tools from statistical mechanics~\cite{mezard_2009,decelle2018thermodynamics,decelle2021restricted}, at the expense of reduced representational capacity.
Various enhancements---Deep Boltzmann Machines~\cite{salakhutdinov_2009}, convolutional RBMs~\cite{lee_2009}, mean-field-inspired augmentations~\cite{mezard_2009}, and reintroducing lateral connections~\cite{shi2019new}---seek to circumvent this issue, but often at the cost of added complexity or computational overhead.

An alternative remedy can be formulated by mimicking neuromodulation in biological systems, where diffusively transmitted chemical signals modulate synaptic plasticity based on the organism's internal state---enabling global coordination beyond the limits of neuronal connectivity~\cite{nadim_2014,shine_2019}. In artificial neural networks, this principle is formalized by three-factor learning rules~\cite{fremaux_2016}, in which a global modulatory signal complements local synaptic updates. Prior implementations along this line have largely employed such signals to dynamically adjust hyperparameters during training~\cite{mei_2023}. However, neuromodulation can also serve to induce competition among units lacking direct connectivity---such as those within an RBM layer---as we demonstrate below.

In this Letter, we propose the Krotov-Hopfield (KH) learning algorithm~\cite{krotov_2019} as a form of neuromodulatory signaling within RBMs.
Although KH was derived from BCM theory~\cite{bienenstock_1982} and shares Hebbian origins~\cite{hebb_1949}, it stands apart by embedding a global competitive mechanism. 
Specifically, the neuron receiving the strongest input in a layer triggers a broad inhibition that discourages other neurons from becoming co-active.
This yields an indirect lateral interaction among hidden units without requiring explicit intralayer connections, which amounts to a rudimentary attention mechanism in a biologically plausible setting that does not rely on backpropagation.  As a result, neurons compete for feature selection rather than patterns competing for neuronal attention. KH algorithm has been used successfully to generate competitive filters for convolutionsl neural nets (CNNs) without employing backpropagation~\cite{grinberg_local_2019}.

In what follows, we demonstrate that KH-modulated RBMs not only improve reconstruction and classification metrics over standard (shallow) RBMs but also exhibit enhanced robustness against weight initialization and overfitting. Thus, we show that
a neuromodulatory-inspired, global inhibition step can effectively offset the absence of lateral interactions to strike
a balance between the expressive modeling power of Boltzmann machines and the computational efficiency of RBMs, resulting in an approach which incorporates the ``semi-Restricted Boltzmann Machine"~\cite{osindero_2007} functionality without significant additional training cost.

\textit{RBM Essentials.---}A shallow RBM consists of binary visible ($\mbf{v}$) and hidden ($\mbf{h}$) units connected via a bipartite graph (Fig.~\ref{fig:graph}) with parameters $\pmb{\theta}=\{\mbf{W},\mbf{a},\mbf{b}\}$.
The energy function is 
\begin{align}
E_{\pmb{\theta}}(\mbf{v},\mbf{h}) \;=\; -\,\mbf{v}^\top \mbf{W}\,\mbf{h} \;-\;\mbf{a}^\top \mbf{v}\;-\;\mbf{b}^\top \mbf{h},
\label{RBM_energy}
\end{align}
and the associated Boltzmann distribution takes the form
$p_{\pmb{\theta}}(\mbf{v,h})=e^{E_{\pmb{\theta}}(\mbf{v,h})}/Z_{\pmb{\theta}}$, 
where $Z_{\pmb{\theta}}= \sum_{\mbf{v,h}}e^{-E_{\mbf{\theta}}(\mbf{v},\mbf{h})}$ is the partition function.
The predicted distribution of the visible units can be obtained by marginalization as $p_{\pmb{\theta}}(\mbf{v})=\sum_{\mbf{h}}p_{\pmb{\theta}}(\mbf{v,h}) \equiv e^{-F_{\pmb{\theta}}(\mbf{v})}/ Z_{\pmb{\theta}}$ and ideally converges to the underlying distribution $p_d (\mbf{v})$ of the dataset upon training. 
$F_{\mbf{\theta}}(\mbf{v})$ is known as the clamped free energy or visible energy.
Accordingly, the average \emph{negative} log-likelihood 
\begin{equation}
\hspace*{-5pt}
    \label{eq:log_l_}
    \mathcal{L} \equiv \langle-\log[p_{\pmb{\theta}}(\mbf{v})]\rangle_{\mbf{v} \sim p_d (\mbf{v})} = \langle F_{\pmb{\theta}}(\mbf{v}) \rangle_{\mbf{v} \sim p_d (\mbf{v})} + \log Z_{\pmb{\theta}}
\end{equation}
serves as a suitable objective function whose minimization by means of gradient descent steps
\begin{align}
    \label{eq:sgd}
    \pmb{\theta}_{t+1} = \pmb{\theta}_{t} - \eta \nabla_{\pmb{\theta}}\mathcal{L}(\pmb{\theta}_t)
\end{align}
can be shown to reduce to
\begin{equation}
    \delta W_{ij} =\eta \left[\,\langle v_i h_j \rangle)_{d} - \langle v_i h_j \rangle_{m}\,\right]
    \label{eq:dW} 
\end{equation}
where $\langle \cdot \rangle_d$ and $\langle \cdot \rangle_m$ denote the expectation values over the data and the model distributions, respectively, and $\eta$ is the learning rate. Updates similar to Eq.\eqref{eq:dW} can be found for the biases $\{\mbf{a},\mbf{b}\}$ as well~\cite{fischer_2012}.

Evaluating $Z_{\pmb{\theta}}$ (therefore the model expectations in Eq.\eqref{eq:dW}) is hard in practice. 
Consequently,
training of RBMs requires approximate techniques which vary from Markov chain Monte Carlo (MCMC) algorithms, such as, contrastive divergence ($\mathrm{CD}$)~\cite{hinton_2002} and persistent $\mathrm{CD}$~\cite{tieleman_2008},  to mean-field-based methods~\cite{goos_2002,marylou_2015}. 
We here use the CD algorithm proposed by Hinton~\cite{hinton_2002}, where an MCMC chain is initiated from a sample drawn from the dataset and iterated with the sequence
\begin{widetext}
\begin{equation}
    \mathbf{v}^{(0)}\sim  p_d(\mathbf{v})\rightarrow \mathbf{h}^{(0)}\sim p_{\pmb{\theta}}(\mathbf{h}^{(0)}|\mathbf{v}^{(0)}) \rightarrow \mathbf{v}^{(1)} \sim p_{\pmb{\theta}}(\mathbf{v}^{(1)}|\mathbf{h}^{(0)}) \cdots \rightarrow 
     \mathbf{h}^{(k-1)}\sim p_{\pmb{\theta}}(\mathbf{h}^{(k-1)}|\mathbf{v}^{(k-1)})\rightarrow    \mathbf{v}^{(k)}\sim p_{\pmb{\theta}}(\mathbf{v}^{(k)}|\mathbf{h}^{((k-1)}). \nonumber 
\end{equation}
\end{widetext}
Calculating the model expectations in Eq.\eqref{eq:dW} using the distributions obtained from the $k$-step Gibbs chain above constitutes the $\mathrm{CD}_k$ algorithm.
It is noteworthy that even $\mathrm{CD_1}$ works reasonably well in practical applications~\cite{hinton_2012}. In order to check the robustness of our findings to the choice of $k$, we ran experiments both with $\mathrm{CD_1}$ and $\mathrm{CD_{10}}$.


\textit{Krotov-Hopfield Updates.---}Krotov and Hopfield's unsupervised learning algorithm~\cite{krotov_2019} incorporates a local, Hebbian-like learning rule for synaptic weight updates.
A synapse $W_{\mu\nu}$ connecting a presynaptic node $\mu$ with activity $x_\mu$ is modified by taking into account the input current to the postsynaptic node $\nu$ given by
\begin{equation}
 I_{r_\nu} = \langle \mbf{W}_{\nu}, \mbf{x}\rangle
 \label{eq:I}
\end{equation}
where $\langle\mathbf{X},\mathbf{Y}\rangle \equiv \sum_{\alpha} X_{\alpha} Y_{\alpha}$  is the Euclidean inner product. For convenience, the components of $\mbf{I}$ are indexed by their rank $r_\nu$ in descending order $I_K \ge I_{K-1}\ge \dots \ge I_1$. The KH weight updates are then expressed as
\begin{equation}
\delta W_{\mu\nu}^{KH} = \varepsilon\,\frac{g_\nu(\mathbf{I})\,\Phi_{\mu\nu}(\mathbf{x},\mathbf{W})}{\max\limits_{\mu\nu}\Bigl[g_\nu(\mathbf{I})\,\Phi_{\mu\nu}(\mathbf{x},\mathbf{W})\Bigr]}
\label{eq:KH_update}
\end{equation}
in terms of the local update function
\begin{equation}
\Phi_{\mu\nu}(\mathbf{x},\mathbf{W}) \equiv R^2 x_{\mu} - \langle \mathbf{W}_{\nu}, \mathbf{x} \rangle\,W_{\mu\nu},
\label{eq:KH_function}
\end{equation}
which reduces to Oja's rule for $g_\nu(\mathbf{I})=\mathbf{I}$.
The first term in Eq.\eqref{eq:KH_function} then amounts to the Hebbian update rule, while the second term is a homeostatic regularization forcing the incoming synaptic weights onto the sphere
$ \sum_{\mu} |W_{\mu\nu}|^2 = R^{2}$
for each postsynaptic unit $\nu$. The hyperparameter $\varepsilon>0$ and the denominator in Eq.\eqref{eq:KH_update} together act as a varying learning rate, which can be interpreted as a context-dependent modulatory effect.

Crucially, Krotov and Hopfield proposed choosing $g_\nu(\mathbf{I})$ as a global modulator~\cite{krotov_2019}
\begin{equation}
g_\nu(\mathbf{I})=\begin{cases}
1, & \text{if} \ r_\nu=K,\\
-\Delta, & \text{if} \ r_\nu=K-\ell, \\
0, & \text{otherwise}.
\end{cases}
\label{eq:KH_modulation}
\end{equation}
where $K$ is number of units in the postsynaptic layer, and $\ell$ and $\Delta>0$ are hyperparameters. We choose $\ell=1$ and $\Delta=0.4$ here. In this scheme, the unit $\nu$ subject to the highest input current ($r_\nu=K$) gets full Hebbian reinforcement, while the next $\ell$ units  ($K>r_\nu \ge K-\ell$) receive an anti-Hebbian feedback where the incoming connections from co-firing presynaptic nodes are {\em weakened} by the factor $\Delta$. The rest of the synaptic weights remain unaltered. $g_\nu(\mathbf{I})$ thus acts as a neuromodulatory signal, inducing a global competition among postsynaptic neurons by selectively applying Hebbian and anti-Hebbian strategies across units without a direct connection.

\textit{KH-modulated RBM.---}Indirect lateral interactions established in the postsynaptic layer by the KH algorithm can offer a computationally efficient approach to enhancing the limited expressive power of RBMs. This can be achieved simply by placing KH updates as a neuromodulatory signaling stage between RBM updates during training (though alternative forms of integration are also conceivable). Specifically, at each time step $t$,
\begin{equation}
\label{eq:mod_preupd}
\pmb{\theta}_t^{KH} =\pmb{\theta}_t +\delta \pmb{\theta}_t^{KH}
\,\text{then}\,\,\,
\pmb{\theta}_{t+1} =\pmb{\theta}_t^{KH} -\eta\,\nabla_{\pmb{\theta}}\mathcal{L}\bigl(\pmb{\theta}_t^{KH}\bigr)
\end{equation}
with the intermediary parameter values $\pmb{\theta}_t^{KH} = \{\mbf{W}_t+\delta \mbf{W}_t^{KH},\mbf{a}_t,\mbf{b}_t\}$ constructed using Eq.\eqref{eq:KH_update}.

We consider two potential implementations which differ in the choice of the input state vector $\mbf{x}$ to Eqs.(\ref{eq:KH_update},\ref{eq:KH_function}):
\begin{enumerate}[label=(\roman*)]
    \item $\mathrm{KH_{TD}}$ (top-down, cognition-driven)  uses $\mbf{h}\sim p_{\pmb{\theta}}(\mbf{h}|\mbf{v})$ for the input $\mbf{x}$ (and $\mbf{W}^\top$ instead of $\mbf{W}$)
    \item $\mathrm{KH_{BU}}$ (bottom-up, sensory-driven) uses $\mathbf{v}\sim p_d(\mbf{v})$ for $\mbf{x}$.
\end{enumerate}
Although both approaches yield qualitatively similar outcomes, we report results for the cognition-driven, top-down modulation, which performed slightly better in our experiments (see Supplemental Material Section B for a detailed comparison). In either case, Eq.\eqref{eq:mod_preupd} can be cast into a Langevin-like form
\begin{align}
    \label{eq:KHsgd}
    \pmb{\theta}_{t+1} = \pmb{\theta}_{t} - \eta\left[\nabla_{\pmb{\theta}}\mathcal{L}(\pmb{\theta}_t) + \pmb{\xi}_t\right]
\end{align}
where 
$ \pmb{\xi}_t\simeq 
\left( \delta \pmb{\theta}_{t}^{KH}\cdot \nabla_{\pmb{\theta}} \right)
\nabla_{\pmb{\theta}}\mathcal{L}(\pmb{\theta}_t)-\eta^{-1}\delta \pmb{\theta}_{t}^{KH}$ 
(obtained by Taylor expansion) encapsulates the total contribution of KH modulation as an additive correction incorporating the local Hessian and the gated Oja's rule in Eq.\eqref{eq:KH_update}. 
\begin{figure}[h!]
\centering
\includegraphics[width=\columnwidth]{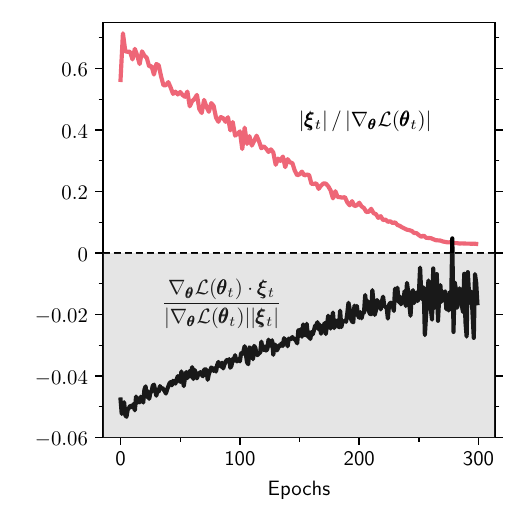}
\caption{\label{fig:comparison} Comparison of the RBM gradient and KH-moduation weight updates during $\mathrm{KH_{TD}}$--RBM training. The red curve shows the time evolution of the magnitude ratio $\bigl|\pmb{\xi}_t\bigr| \,/\, \bigl|\nabla_{\pmb{\theta}}\mathcal{L}\bigl(\pmb{\theta}_t\bigr)\bigr|$ from Eq.~\eqref{eq:KHsgd}. The black curve is the directional correlation measured by the cosine similarity, 
$\nabla_{\pmb{\theta}}\mathcal{L}\bigl(\pmb{\theta}_t\bigr)\cdot\pmb{\xi}_t / \bigl|\nabla_{\pmb{\theta}}\mathcal{L}\bigl(\pmb{\theta}_t\bigr)\bigr|\,\bigl|\pmb{\xi}_t\bigr|$.  KH modulation is applied with a time-dependent (scheduled) learning rate and is fully phased out at 300 epochs. Each data point is an average of over 20 time steps, and each time step is averaged over 3000 trials.}
\end{figure}

Upon experimentation, we found that KH updates perform best in an ``annealed" setting, where the KH step size ($\varepsilon$) is gradually reduced according to a predefined schedule, as detailed below. A comparison, in terms of the magnitudes and the directions, of the RBM updates, $\nabla_{\pmb{\theta}}\mathcal{L}(\pmb{\theta}_t)$, and KH modulation, $\pmb{\xi}_t$, in Eq.\eqref{eq:KHsgd} during a typical training run is given in Fig.~\ref{fig:comparison}. While the decay in the magnitude ratio is by design (the annealing protocol), the existence of a negative directional correlation between the RBM updates and the KH contribution is surprising. 
Upon inspecting Fig.~\ref{fig:comparison}, the KH modulation appears to direct the search primarily perpendicular to the gradient (as a random vector would in high dimensions), while slightly backtracking along the calculated RBM gradient.

In view of the above observations, one might wonder if the additive contribution of KH in Eq.\eqref{eq:KHsgd} ultimately plays a role similar to noise injection in the ``Stochastic Gradient Langeving Dynamics" (SGLD) method~\cite{sgld}, which can facilitate better posterior sampling and mitigate overfitting. To test this possibility, we have checked that replacing the KH contribution by a random vector (obtained by shuffling the components of $\pmb{\xi}_t$) does not yield the performance boost detailed below (see Supplemental Material Section A). Therefore, rather than acting as a random perturbation, KH modulation appears to systematically steer the RBM training trajectory
toward an alternative minimum on the landscape where a competition-driven weight assignment allows better use of available capacity. Accordingly, we found that hidden-unit receptive fields, $\mathbf{W}_\nu$, exhibit reduced overlap when KH modulation is applied. In particular, the average cosine similarity between a receptive field and its maximally overlapping alternative is 0.41 for shallow RBMs, compared to 0.37 for $\mathrm{KH_{TD}}$--RBM and 0.35 $\mathrm{KH_{BU}}$--RBM models after training.

We next demonstrate that KH-modulated RBM improves on the standard version in terms of the validation accuracy (or achieves on-par performance with a fraction of the number of parameters) while simultaneously eliminating overfitting in two classical machine learning tasks, reconstruction and classification. 
To this end, we conducted a series of experiments on the Modified National Institute of Standards and Technology (MNIST)~\cite{mnist} dataset whose 70000 samples were partitioned it into 60000 training samples and 10000 validation samples. Each sample is a $28 \times 28$ binary image obtained by thresholding the pixel values $v_i \in [0,255]$ of the original data using a global threshold of 127. The dataset---being images of handwritten digits---consists of ten classes with a balanced representation in both the training and the validation sets which are otherwise random.

We used two RBMs composed of $M=100$ and $M=500$ hidden units, and conducted each experiment on both sizes. The models were trained using Contrastive Divergence, $\mathrm{CD}_k$, with $k=1$ and $k=10$ Gibbs sampling steps, separately. The learning rate and the batch size for all RBMs were set to $\eta=0.1$ and $n_b = 100$, respectively, regardless of whether a KH-RBM or a shallow RBM was trained. 

\begin{figure}[t]
\centering
\includegraphics[width=\columnwidth]{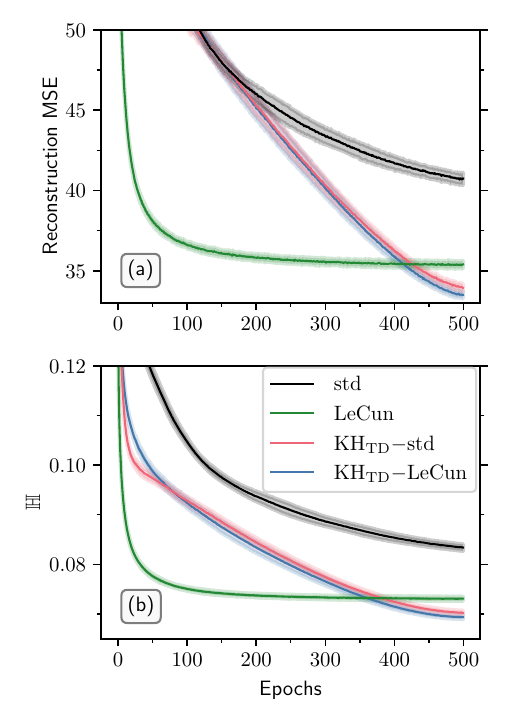}
\caption{\label{fig:td_fig}Comparison of $\mathrm{KH_{TD}}$--RBMs with the shallow RBMs, each with $\mathrm{CD_1}$ and $M=100$: (a) Reconstruction MSE, (b) cross entropy. The final stage of the training where the improved performance of KH-modulation over standard RBM on the validation dataset is evident even with $\mathrm{CD_1}$.}
\end{figure}

Weight initialization can make a significant difference in the training performance of standard RBMs. 
We therefore report the validation performance for two different and commonly used initializations~\cite{LeCun_2012}, namely, StdNormal: $W_{\mu\nu}\sim\mathcal{N}(0,1)$ and LeCun: $W_{\mu\nu}\sim\mathcal{N}\left(0, 1 / N\right)$, where $\mathcal{N}(0,\sigma^2)$ signifies the normal distribution with zero mean and a standard deviation $\sigma$. The step-size parameter of the KH algorithm was set to decay as $\varepsilon = \varepsilon_0 \left(1 - n_\mathrm{epoch}/S\right)^{3/2}$ in which $n_\mathrm{epoch}$ is the current epoch number and $S$ is the total number of epochs. This scheduling sets a finite time scale for KH modulation. The power $3/2$ was chosen by trial and error, with the intuition that a super-linear decay provides sufficient time for a (almost) pure RBM gradient descent to converge to a minimum in the region targeted by KH modulation. In all experiments, the initial step size $\varepsilon_0$ and the penalty parameter $\Delta$  in Eq.\eqref{eq:KH_modulation} were chosen optimally by means of a grid search over a hyperparameter set, for a fair comparison. Finally, we chose $R^{(\mathrm{std})}_\nu=1$  and $R^{(\mathrm{LeCun})}_\nu=0.1$ for all $\nu$, making sure that the synaptic weights for a post-synaptic unit do not accidentally land on the fixed sphere of Eq.\eqref{eq:KH_update} (yielding vanishing KH updates) ~\footnote{The code implementations are made available on GitHub: \url{https://github.com/basertambas/KH-RBM}.}.

In Fig.~\ref{fig:td_fig}, we compare the training progression of $\mathrm{KH_{TD}}$--$\mathrm{CD_1}$ and shallow RBMs with $M=100$, by monitoring the difference between the predicted and the true distribution of the validation data in terms of both the reconstruction mean-squared error
\begin{align}
   \text{MSE} = \frac{1}{N}\sum_{i=1}^{N}(v_i - \hat{v}_{i})^{2}, 
\end{align}
and the cross entropy
\begin{eqnarray}
    \mathbb{H}&=&-\sum_{i=1}^{N} v_i \log(p_{\pmb{\theta}}(\hat{v}_i=1|\mathbf{h})) \nonumber \\
    && +(1-v_i) \log(1-p_{\pmb{\theta}}(\hat{v}_i=1|\mathbf{h})).
\end{eqnarray}
 Reported quantities are averages over ten independent training runs with different initial random seeds.

Several aspects of $\mathrm{KH_{TD}}$--RBM stand out in Fig.~\ref{fig:td_fig}: (i) training is robust to the choice of initialization, unlike the standard RBM, (ii) the convergence rate is slow compared to LeCun-initialized standard RBMs but is better than StdNormal's, (iii) for both initializations, the final trained network outperforms the standard RBMs on the validation dataset.

\begin{figure}[t!]
\centering
\includegraphics[width=\columnwidth]{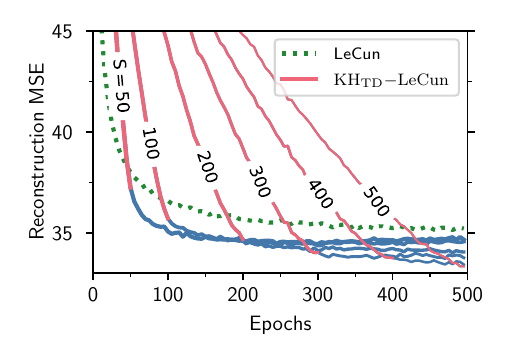}
\caption{\label{fig:m_sch_fig}Reconstruction MSE on the validation dataset during the training of 
$\mathrm{KH_{TD}}$--RBMs with the MNIST dataset for 500 epochs. We show the average of 10 independent runs using $M=100$, $\mathrm{LeCun}$ initialization, $\mathrm{CD_1}$, and different durations ($S$) of the scheduling protocol (shown in red).}
\end{figure}

In Fig.~\ref{fig:td_fig}, the high convergence rate of shallow RBMs initialized with LeCun's scheme stands out as a key advantage over KH-RBMs, as faster training may be more desirable than marginally better accuracy in some applications. We therefore investigated whether KH-modulation could be adapted to strike a more favorable balance between speed and performance. We found that it is possible to achieve an almost equally fast convergence to a better validation MSE simply by decaying the KH-modulation updates more rapidly.  In Fig.~\ref{fig:m_sch_fig} we report the validation MSE for $\mathrm{KH_{TD}}$--RBMs trained with the scheduling hyperparameter $S\in \{50,100,200,300,400,500\}$, such that, once the KH modulation fully decays (after $S$ epochs) pure RBM update inherits the dynamics. In all cases, $\mathrm{KH_{TD}}$--RBM outperformed shallow RBM at the end of the KH modulation window and remained so during the rest of the training, irrespective of how small $S$ is.

\begin{figure}[t]
\centering
\includegraphics[width=\columnwidth]{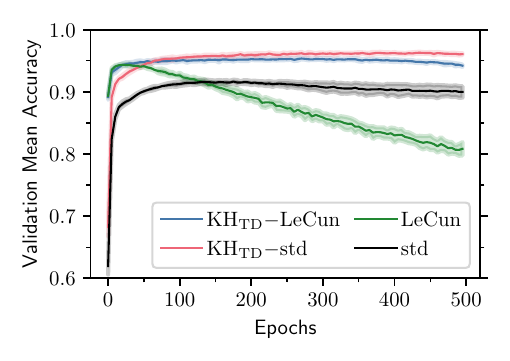}
\caption{\label{fig:m_acc_fig}Prediction accuracy {\em vs} epochs calculated on the validation dataset for the MNIST classification task. The figure compares of $\mathrm{KH-cRBMs}$ with shallow cRBMs, both using $\mathrm{CD_1}$ and $M=500$.}
\end{figure}

In order to test whether KH modulation has a positive impact also on a more traditional machine-learning task, we next applied it to cRBMs~\cite{larochelle_2012} for MNIST digit classification. To this end, we used graph structures with $M=100$ and $M=500$ units with the two different weight initializations, and $k=1$ and $k=10$ Gibbs sampling steps, as before. The data presented in Figure~\ref{fig:m_acc_fig} and Table~\ref{tab:m_acc} indicate that KH modulation not only improves the validation accuracy of the standard cRBM but also nearly eliminates overfitting---a benefit that is most pronounced in the LeCun-initialized training runs. With $500$ hidden units, StdNormal initialization and 10 Gibbs sampling steps (CD$_{10}$), the network reaches its maximum accuracy of $\%97$. This value is slightly reduced with 1-step Gibbs sampling to $\%96.5$, which is the accuracy reported for a shallow cRBM trained with 6000 hidden units~\cite{larochelle_2012}. In other words, KH modulation allows a 10-fold reduction in model size without sacrificing from accuracy. Alternatively, comparing models of identical size in Table~\ref{tab:m_acc}, KH modulation yields consistent gains in validation accuracy, with improvements exceeding 1\% in accuracy (more than $25\%$ reduction in error rate) for $M=500$. We note that, although higher accuracies have been reported for MNIST classification using specialized RBM variants, these results were derived from a generic implementation, without classification-specific tuning or architectural optimizations.

\begin{table}[tbp]
\caption{\label{tab:m_acc}The maximum validation accuracies for the MNIST classification task achieved using both the proposed and standard cRBM implementations. Accuracies are reported as a function of model size and initialization method, with the best performance for each model size highlighted in bold. Training runtimes per epoch are also given for each model.}
\centering
\begin{ruledtabular}
\scalebox{1.0}{
\begin{tabular}{lcccccr}
\multicolumn{1}{c}{}    & \multicolumn{2}{l}{$M=100$}        &           & \multicolumn{2}{l}{$M=500$}                     \\ 
\cline{2-7}
\multicolumn{1}{l}{}    & $\mathrm{LeCun}$ & $\mathrm{std}$ & T(s) & $\mathrm{LeCun}$ & $\mathrm{std}$ &T(s) \\ 
\hline\hline
$\mathrm{CD_{1}}$                & 90.6(2)               & 91.3(2)     &1.1        & 94.6(1)                & 92.0(3)     &2.5         \\ 
$\mathrm{KH_{TD}}$--$\mathrm{CD_1}$  & 91.2(3)                & \bf{92.1(3)}   &2.2          & 95.6(1)                & 96.5(1)  & 4.5            \\ 
\hline
$\mathrm{CD_{10}}$              & 90.4(3)               & 90.8(2)    &3.9         & 95.8(3)                & 93.8(2)     &9.0         \\
$\mathrm{KH_{TD}}$--$\mathrm{CD_{10}}$ & 90.5(3)               & 91.2(2)  &5.0           & 96.9(1)                & \bf{97.0(1)}   &11.1            \\ 
\end{tabular}
}
\end{ruledtabular}
\end{table}

All the experiments described above, including reconstruction and classification tasks, were repeated for the Kuzushiji MNIST~\cite{kmnist} (KMNIST) dataset, which is considered to be a more challenging version of MNIST. The results reported in Supplemental Material Section D demonstrate a similar performance boost, increasing the confidence on the validity of our observations across applications.

In conclusion, we presented a proof-of-concept implementation of the KH algorithm as a form of neuromodulatory signaling in RBMs. We showed that the selective inhibition of synaptic updates proposed by Krotov and Hopfield---which can be interpreted as a form of cognition-driven (TD) or sensory-driven (BU) attention mechanism without backpropagation--- leads to visible improvements in both reconstruction and classification tasks. Furthermore, KH-modulated RBMs exhibit reduced sensitivity to weight initialization and a marked resilience against overfitting.
Typically, the absence of lateral connections in the bipartite structure of RBMs can result in redundant feature learning due to the well-known issue of hidden unit co-adaptation~\cite{hinton_coad}. Incorporating the KH algorithm into the RMB training process appears to partially overcome this problem by promoting more diverse representations through a global competition scheme, thereby enhancing the network’s capacity without need for significant additional resources.

A detailed comparison of the minima reached with and without KH modulation is planned as future work and should allow a better understanding of the gains reported above. Finally, the performance of KH-RBMs encourages the use of a similar, hybrid framework in more complex architectures
which may further bridge the gap between biologically plausible learning and state-of-the-art machine learning approaches.


\begin{acknowledgments}
We acknowledge beneficial discussions with A.T. Y\i ld\i r\i m and D. Yuret, as well as a partial support by the Technological and Scientific Research Council of T\" urkiye (T\" UB\. ITAK) through the grant MFAG-119F121 during the initial stages of this work.
\end{acknowledgments}

\bibliography{bibliography}

\end{document}